\documentclass[prl,twocolumn,superscriptaddress,amsmath,amssymb]{revtex4}

\usepackage{color}
\usepackage{times}
\usepackage{fancyhdr}
\usepackage{float}
\usepackage{alltt}
\usepackage{listings}
\usepackage{enumerate}
\usepackage[T1]{fontenc}
\usepackage{ae}
\usepackage{subfigure}
\usepackage{amsmath}
\usepackage{amssymb}
\usepackage{graphicx}
\usepackage{dcolumn}
\usepackage{bm}

\def\be{\begin{equation}}
\def\ee{\end{equation}}
\def\beg{\begin{align}}
\def\eeg{\end{align}}
\def\bi{\begin{itemize}}
\def\ei{\end{itemize}}
\def\ben{\begin{enumerate}[1.]}
\def\een{\end{enumerate}}

\newcommand{\bo}{\raise-1mm\hbox{\Large$\Box$}}
\newcommand{\sci}[2]{#1 \times 10^{#2}}
\newcommand{\rthz}{\mathrm{Hz}^{-\frac{1}{2}}}
\newcommand{\hrss}{h_{\mathrm{rss}}}

\newcommand{\hrssn}{h_{\mathrm{rss}}^{90\%}}

\newcommand{\egwn}{E_{\mathrm{GW}}^{90\%}}
\newcommand{\egw}{E_{\mathrm{GW}}}
\newcommand{\eem}{E_{\mathrm{EM}}}

\begin{document}

\title{Search for Gravitational Wave Bursts from Soft Gamma Repeaters}

\newcommand*{\AG}{Albert-Einstein-Institut, Max-Planck-Institut f\"ur Gravitationsphysik, D-14476 Golm, Germany}
\affiliation{\AG}
\newcommand*{\AH}{Albert-Einstein-Institut, Max-Planck-Institut f\"ur Gravitationsphysik, D-30167 Hannover, Germany}
\affiliation{\AH}
\newcommand*{\AU}{Andrews University, Berrien Springs, MI 49104 USA}
\affiliation{\AU}
\newcommand*{\AN}{Australian National University, Canberra, 0200, Australia}
\affiliation{\AN}
\newcommand*{\CH}{California Institute of Technology, Pasadena, CA  91125, USA}
\affiliation{\CH}
\newcommand*{\CA}{Caltech-CaRT, Pasadena, CA  91125, USA}
\affiliation{\CA}
\newcommand*{\CU}{Cardiff University, Cardiff, CF24 3AA, United Kingdom}
\affiliation{\CU}
\newcommand*{\CL}{Carleton College, Northfield, MN  55057, USA}
\affiliation{\CL}
\newcommand*{\CS}{Charles Sturt University, Wagga Wagga, NSW 2678, Australia}
\affiliation{\CS}
\newcommand*{\CO}{Columbia University, New York, NY  10027, USA}
\affiliation{\CO}
\newcommand*{\ER}{Embry-Riddle Aeronautical University, Prescott, AZ   86301 USA}
\affiliation{\ER}
\newcommand*{\HC}{Hobart and William Smith Colleges, Geneva, NY  14456, USA}
\affiliation{\HC}
\newcommand*{\IA}{Institute of Applied Physics, Nizhny Novgorod, 603950, Russia}
\affiliation{\IA}
\newcommand*{\IU}{Inter-University Centre for Astronomy  and Astrophysics, Pune - 411007, India}
\affiliation{\IU}
\newcommand*{\HU}{Leibniz Universit{\"a}t Hannover, D-30167 Hannover, Germany}
\affiliation{\HU}
\newcommand*{\CT}{LIGO - California Institute of Technology, Pasadena, CA  91125, USA}
\affiliation{\CT}
\newcommand*{\LM}{LIGO - Massachusetts Institute of Technology, Cambridge, MA 02139, USA}
\affiliation{\LM}
\newcommand*{\LO}{LIGO Hanford Observatory, Richland, WA  99352, USA}
\affiliation{\LO}
\newcommand*{\LV}{LIGO Livingston Observatory, Livingston, LA  70754, USA}
\affiliation{\LV}
\newcommand*{\LU}{Louisiana State University, Baton Rouge, LA  70803, USA}
\affiliation{\LU}
\newcommand*{\LE}{Louisiana Tech University, Ruston, LA  71272, USA}
\affiliation{\LE}
\newcommand*{\LL}{Loyola University, New Orleans, LA 70118, USA}
\affiliation{\LL}
\newcommand*{\MS}{Moscow State University, Moscow, 119992, Russia}
\affiliation{\MS}
\newcommand*{\ND}{NASA/Goddard Space Flight Center, Greenbelt, MD  20771, USA}
\affiliation{\ND}
\newcommand*{\NA}{National Astronomical Observatory of Japan, Tokyo  181-8588, Japan}
\affiliation{\NA}
\newcommand*{\NO}{Northwestern University, Evanston, IL  60208, USA}
\affiliation{\NO}
\newcommand*{\RA}{Rutherford Appleton Laboratory, Chilton, Didcot, Oxon OX11 0QX United Kingdom}
\affiliation{\RA}
\newcommand*{\SJ}{San Jose State University, San Jose, CA 95192, USA}
\affiliation{\SJ}
\newcommand*{\SM}{Sonoma State University, Rohnert Park, CA 94928, USA}
\affiliation{\SM}
\newcommand*{\SE}{Southeastern Louisiana University, Hammond, LA  70402, USA}
\affiliation{\SE}
\newcommand*{\SO}{Southern University and A\&M College, Baton Rouge, LA  70813, USA}
\affiliation{\SO}
\newcommand*{\SA}{Stanford University, Stanford, CA  94305, USA}
\affiliation{\SA}
\newcommand*{\SR}{Syracuse University, Syracuse, NY  13244, USA}
\affiliation{\SR}
\newcommand*{\PU}{The Pennsylvania State University, University Park, PA  16802, USA}
\affiliation{\PU}
\newcommand*{\TA}{The University of Texas at Austin, Austin, TX 78712, USA}
\affiliation{\TA}
\newcommand*{\TC}{The University of Texas at Brownsville and Texas Southmost College, Brownsville, TX  78520, USA}
\affiliation{\TC}
\newcommand*{\TR}{Trinity University, San Antonio, TX  78212, USA}
\affiliation{\TR}
\newcommand*{\BB}{Universitat de les Illes Balears, E-07122 Palma de Mallorca, Spain}
\affiliation{\BB}
\newcommand*{\UA}{University of Adelaide, Adelaide, SA 5005, Australia}
\affiliation{\UA}
\newcommand*{\BR}{University of Birmingham, Birmingham, B15 2TT, United Kingdom}
\affiliation{\BR}
\newcommand*{\FA}{University of Florida, Gainesville, FL  32611, USA}
\affiliation{\FA}
\newcommand*{\GU}{University of Glasgow, Glasgow, G12 8QQ, United Kingdom}
\affiliation{\GU}
\newcommand*{\MD}{University of Maryland, College Park, MD 20742 USA}
\affiliation{\MD}
\newcommand*{\MA}{University of Massachusetts, Amherst, MA 01003 USA}
\affiliation{\MA}
\newcommand*{\MU}{University of Michigan, Ann Arbor, MI  48109, USA}
\affiliation{\MU}
\newcommand*{\MN}{University of Minnesota, Minneapolis, MN 55455, USA}
\affiliation{\MN}
\newcommand*{\OU}{University of Oregon, Eugene, OR  97403, USA}
\affiliation{\OU}
\newcommand*{\RO}{University of Rochester, Rochester, NY  14627, USA}
\affiliation{\RO}
\newcommand*{\SL}{University of Salerno, 84084 Fisciano (Salerno), Italy}
\affiliation{\SL}
\newcommand*{\SN}{University of Sannio at Benevento, I-82100 Benevento, Italy}
\affiliation{\SN}
\newcommand*{\SH}{University of Southampton, Southampton, SO17 1BJ, United Kingdom}
\affiliation{\SH}
\newcommand*{\SC}{University of Strathclyde, Glasgow, G1 1XQ, United Kingdom}
\affiliation{\SC}
\newcommand*{\WA}{University of Western Australia, Crawley, WA 6009, Australia}
\affiliation{\WA}
\newcommand*{\UW}{University of Wisconsin-Milwaukee, Milwaukee, WI  53201, USA}
\affiliation{\UW}
\newcommand*{\WU}{Washington State University, Pullman, WA 99164, USA}
\affiliation{\WU}

\author{B.~Abbott}    \affiliation{\CT}
\author{R.~Abbott}    \affiliation{\CT}
\author{R.~Adhikari}    \affiliation{\CT}
\author{P.~Ajith}    \affiliation{\AH}
\author{B.~Allen}    \affiliation{\AH}  \affiliation{\UW}
\author{G.~Allen}    \affiliation{\SA}
\author{R.~Amin}    \affiliation{\LU}
\author{S.~B.~Anderson}    \affiliation{\CT}
\author{W.~G.~Anderson}    \affiliation{\UW}
\author{M.~A.~Arain}    \affiliation{\FA}
% ...

 \collaboration{The LIGO Scientific Collaboration, http://www.ligo.org}
 \noaffiliation

\newcommand*{\IASF}{IASF Milano via E.Bassini 15, I-20133 Milano,
Italy}
\newcommand*{\BKSSL}{University of California-Berkeley, Space Sciences
Lab, 7 Gauss Way, Berkeley, CA 94720, USA}
\newcommand*{\ALAMOS}{Los Alamos National Laboratory, Los Alamos, NM, 87545, USA}

\author{S.~Barthelmy} \affiliation{\ND}
\author{N.~Gehrels} \affiliation{\ND}
\author{K. C.~Hurley} \affiliation{\BKSSL}
\author{D.~Palmer} \affiliation{\ALAMOS}

\begin{abstract}

We present the results of a LIGO search for short-duration
gravitational waves (GWs) associated with Soft Gamma Repeater (SGR)
bursts.  This is the first search sensitive to neutron star
$f$-modes, usually considered the most efficient GW emitting modes.
We find no evidence of GWs associated with any SGR burst in a sample
consisting of the 27 Dec.\ 2004 giant flare from SGR 1806$-20$ and
190 lesser events from SGR 1806$-20$ and SGR 1900+14 which occurred
during the first year of LIGO's fifth science run. GW strain upper
limits and model-dependent GW emission energy upper limits are
estimated for individual bursts using a variety of simulated
waveforms. The unprecedented sensitivity of the detectors allows us
to set the most stringent limits on transient GW amplitudes
published to date. We find upper limit estimates on the
model-dependent isotropic GW emission energies (at a nominal
distance of 10\,kpc) between $\sci{3}{45}$ and $\sci{9}{52}$ erg
depending on waveform type, detector antenna factors and noise
characteristics at the time of the burst.  These upper limits are
within the theoretically predicted range of some SGR models.

\end{abstract}

\pacs{
04.80.Nn,
07.05.Kf
95.85.Sz
97.60.Jd
 }

\maketitle

Soft Gamma Repeaters (SGRs) sporadically emit brief
($\approx0.1$\,s) intense bursts of soft gamma-rays with peak
luminosities commonly up to $10^{42}$ erg/s \,\cite{mereghetti08,
woods04a}. Less common intermediate bursts with greater peak
luminosities can last for seconds.  Rare ``giant flare'' events,
some 1000 times brighter than common bursts\,\cite{palmer05}, have
initial bright, short ($\approx0.2$\,s) pulses followed by tails
lasting minutes and are among the most electromagnetically luminous
events in the Universe\,\cite{woods04a}. Since the discovery of SGRs
in 1979\, three of the four confirmed SGRs
have produced a giant flare each~\cite{mazets79,hurley99,hurley05}.

SGRs are promising sources of gravitational waves (GWs).
According to the ``magnetar'' model SGRs are neutron stars with
exceptionally strong magnetic fields
$\sim\nolinebreak10^{15}$\,G\,\cite{duncan92}.  SGR bursts may
result from the interaction of the star's magnetic field with its
solid crust, leading to crustal deformations and occasional
catastrophic cracking\,\cite{thompson95, schwartz05} with subsequent
excitation of the star's nonradial modes\,\cite{andersson97,
pacheco98, ioka01} and the emission of GWs\,\cite{horvath05, pacheco98, ioka01}.  Excitation of nonradial
modes could also occur if SGRs are instead solid quark
stars\,\cite{xu03, owen05, horvath05}.

We present a search for short-duration  GW signals
($\lesssim0.3$\,s) associated with SGR bursts using data collected
by the Laser Interferometer Gravitational Wave Observatory (LIGO)
\,\cite{S5}. LIGO consists of two co-located GW detectors at
Hanford, WA with baselines of 4\,km and 2\,km and one 4\,km detector
at Livingston, LA. GW data from one or two of these detectors are
used. When three detectors are operating, data from the most
sensitive pair are chosen.

The SGR burst sample was provided by gamma-ray satellites of the
interplanetary network\,\cite{ipn}, and includes the 27 Dec.\ 2004
giant flare from SGR 1806$-20$ and 214 confirmed bursts (152 from
SGR 1806$-20$ and 62 from SGR 1900+14, one of which was a
multi-episodic ``storm''\,\cite{israel08}) occurring during the
first year of LIGO's fifth science run (S5) from 14 Nov.\ 2005 to 14
Nov.\ 2006. Of the 214 bursts, 117 occurred with three LIGO
detectors operating, 53 with two detectors operating, 20 with a
single detector operating, and 24 with no detector operating.
Including the giant flare, analysis was possible for a total of 191
listed SGR events.

To analyze a given SGR burst we divide the GW data into an on-source
time region, in which GWs associated with the burst could be
expected, and a background time region. In the background region we
do not expect a GW associated with the SGR burst, but the noise is
statistically similar to the on-source region. For isolated bursts
the on-source region consists of 4\,s of data centered on the SGR
burst. GW emission is expected to occur almost simultaneously with
the electromagnetic burst\,\cite{ioka01}; the 4\,s on-source
duration accounts for uncertainties in the geocentric
electromagnetic peak time. There are three special cases: 1) for two
SGR 1900+14 bursts which occurred within 4\,s of each other a
combined 7\,s on-source region was chosen; 2) for the SGR 1900+14
storm a 40\,s on-source region was used; 3) for the GRB 060806 event
from SGR 1806$-20$, two 4\,s on-source regions were used, centered
on the two distinct bright bursts comprising the event.
Background regions consist of 1000\,s of good data on either side of
on-source regions.
On-source and background segments are analyzed identically,
including data quality cuts, resulting in lists of ``analysis
events.'' Analysis events from the background regions are used to
estimate the significance of the on-source analysis events;
significant events, if any, are subject to environmental vetoes and
consistency checks.

The analysis is performed by the \emph{Flare
pipeline}\,\cite{kalmus07, kalmus08, flareCVS} and is based on the
excess power detection statistic of~\cite{anderson01}. Search
parameters such as frequency bands and time windows
are chosen to optimally detect the target signals.  This is achieved
by comparing detection efficiencies for simulated target signals
injected into the background data and searched for with different
search parameters\,\cite{kalmus08}.
The search targets neutron star fundamental mode
ringdowns (RDs) predicted in\,\cite{thorne83,
andersson97, pacheco98, ioka01, andersson02} as well as unmodeled
short-duration GW signals. Model predictions from\,\cite{benhar04}
for ten realistic neutron star equations of state give $f$-mode RD
frequencies in the range 1.5--3\,kHz and damping times in the range
100--400\,ms. We use a search band 1--3\,kHz for RD searches (to
include stiffer equations of state), and find a 250\,ms time window
to be optimal.  The search for unmodeled signals uses time windows
set by prompt SGR burst timescales (5--200\,ms) and frequency bands
set by the detector's sensitivity; a 125\,ms time window
effectively covers this duration range, and we search in two
bands: 100--200\,Hz (probing the region in which the detectors are
most sensitive) and 100--1000\,Hz (for full spectral coverage below the ringdown search band).

In the absence of a detection, for each SGR burst we estimate
loudest event upper limits\,\cite{brady04} on the GW strain incident
on the detector, $h_{\mathrm{rss}}$. Following \cite{s2burst}
$h_{\mathrm{rss}}^{2} =
 h_{\mathrm{rss+}}^{2} + h_{\mathrm{rss\times}}^{2}$, where e.g.
$h_{\mathrm{rss+}}^{2} = \int_{-\infty}^{\infty} h_{+}^{2} dt$ and
$h_{\mathrm{+,\times}}(t)$
are the two GW polarizations. The relationship between the GW polarizations and the
detector response $h(t)$ to an impinging GW from an altitude and azimuth
$(\theta,\phi)$ and with polarization angle $\psi$ is:
 \be
    h(t) = F^{+}(\theta, \phi, \psi) h_+(t)   +  F^{\times}(\theta, \phi, \psi)
 h_{\times}(t)
 \label{eq:hsim}
 \ee
where $F^{+}(\theta, \phi, \psi)$ and $F^{\times}(\theta, \phi,
\psi)$ are the antenna functions for the source at
$(\theta,\phi)$\,\cite{300years}. The upper limit is computed in a
frequentist framework following the commonly used procedure of
injecting simulated signals in the data and recovering them using
the search pipeline\,(see for example \cite{S2inspiral,S2S3S4GRB}).
The upper limits are derived for RD signals and for unmodeled bursts. Correspondingly, RD and band-limited white noise burst (WNB) waveforms are injected with parameters chosen to probe the respective target signal space. For
WNBs independent polarization components are generated
with $h_{\mathrm{rss}+} = h_{\mathrm{rss}\times}$. For RD signals linearly and circularly polarized waves are considered.

The GW strain $\hrss$ upper limits can be recast as upper limits
on the emitted GW energy, $E_{\mathrm{GW}}$. Assuming isotropic
emission, the GW energy associated with $h_{+}(t)$ and
${h}_{\times}(t)$ is~\cite{shapiro83}: \be E_{\mathrm{GW}} = 4\pi
R^2 \frac{c^3}{16 \pi G} \int_{-\infty}^{\infty}\left(
(\dot{h}_{+})^2 + (\dot{h}_{\times})^2\right) dt. \ee We use this
equation with a nominal source distance of $R=10$\,kpc (source
locations and distances are discussed in\,\cite{corbel04,
kaplan02b}) to compute the energies associated with the $\hrss$
upper limits for different signals.

\emph{Results --- }
 We find no evidence for
gravitational waves associated with any of the SGR burst events in
the sample. The significance of on-source analysis events is inferred by
assigning rates at which background analysis events of equal or
greater loudness occur.  We find the most significant on-source
analysis event occurs at a rate of $\sci{1.35}{-3}$\,Hz (1 per
741\,s), which is consistent with the expectation for the 803\,s of
on-source data in the sample. We estimate 90\% confidence strain and energy upper
limits, $\hrssn$ and $\egwn$, using the loudest on-source analysis
event for each SGR burst. Upper limits depend on detector
sensitivity and antenna factors at the time of the burst, the
loudest on-source analysis event, and the simulation waveform type
used.

Table\,\ref{table:bestresults} lists upper limits for the SGR
1806$-20$ giant flare and for the brightest peak of the GRB 060806
event\,\cite{gcn5426} (complete results are given in\,\cite{epaps}).
Results from these two events are highlighted because they yield the
smallest values of $\gamma = \egwn/E_{\mathrm{EM}}$, a measure of
the extent to which an energy upper limit probes the GW emission
efficiency. At the time of the giant flare the LIGO Hanford 4\,km
detector was operating during a commissioning period (LIGO
Astrowatch) and had noise amplitude higher than that of S5 by a
factor of 3; the rms antenna factor, which is an indicator of the
average sensitivity to a given source in the sky, for such event was
$(F_{+}^2 + F_{\times}^2)^{1/2}=0.3$.  The isotropic electromagnetic
energy ($\eem$) for the event, assuming a distance of 10\,kpc, was
$\sci{1.6}{46}$\,erg\,\cite{hurley05}.  At the time of GRB 060806
both the 4\,km and 2\,km Hanford detectors
were observing, with rms antenna factor for that event of 0.5.
$\eem$ for the brightest peak of GRB 060806 was at least
$\sci{2.9}{42}$\,erg\,\cite{gcn5426}.

We estimate upper limits on GW strain and isotropic GW energy
emitted using RDs with $\tau = 200$\,ms at various frequencies, and
WNBs lasting 11 and 100\,ms and with  100--200 and 100--1000\,Hz
bands. We observe no more than 15\% degradation in strain upper
limits using RDs with $\tau$ in the range 100--300\,ms, and no more
than 20\% degradation using WNBs with durations in the range
5--200\,ms, as compared to the upper limits obtained for the nominal
RDs and WNBs used for tuning the search.
Superscripts in Table\,\ref{table:bestresults} give a
systematic error and uncertainties at 90\% confidence. The first and
second superscripts account for systematic error and statistical
uncertainty in amplitude and phase of the detector calibrations,
estimated via Monte Carlo simulations, respectively. The third is a
statistical uncertainty arising from using a finite number of
injected simulations, estimated with the bootstrap method using 200
ensembles\,\cite{efron79}.  The systematic error and the quadrature
sum of the statistical uncertainties are added to the final upper
limit estimates.

\begin{figure}[!t]
\begin{center}
\includegraphics[angle=0,width=90mm, clip=false]{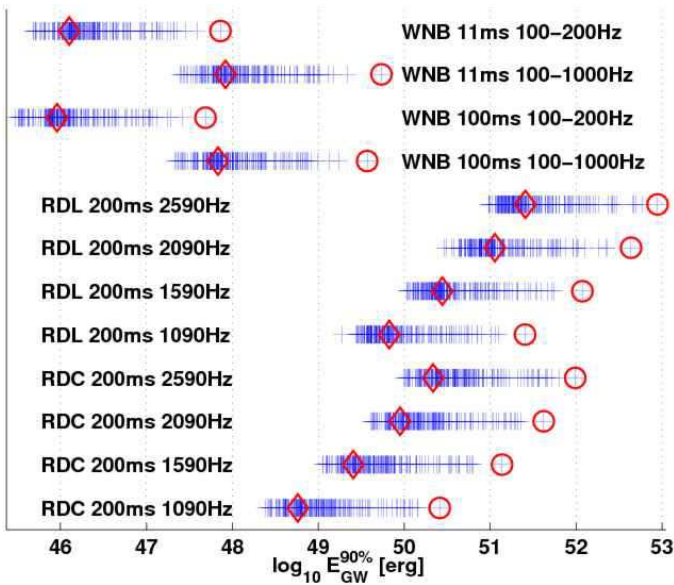}
\caption{ $\egwn$ upper limits for the entire SGR burst sample
for various circularly/linearly polarized RD (RDC/RDL) and
white noise burst (WNB) signals.  The limits shown in
Table\,\ref{table:bestresults}, for the giant flare and GRB 060806,
are indicated in the figure by circles and diamonds, respectively.}
\label{fig:hrss90histograms}
\end{center}
\end{figure}

Figure\,\ref{fig:hrss90histograms} shows $\egwn$ limits for the
entire SGR burst sample.  The lowest upper limit in the sample,
$\egwn=\sci{2.9}{45}$\,erg, is obtained for an SGR 1806$-20$ burst
on 21 Jul. 2006, with a geocentric crossing time of 17:10:56.6 UTC.
The lowest upper limit from the RD search is
$\egwn=\sci{2.4}{48}$\,erg for an SGR 1806$-20$ burst on 24 Aug.
2006 14:55:26 UTC.

\emph{Discussion --- }Two searches for GWs associated with SGR
events have been published previously; neither claimed detection.  The
AURIGA collaboration searched for GW bursts associated with the SGR
1806$-20$ giant flare in the band 850--950~Hz with damping time
100~ms, setting upper limits on the GW energy of
$\sim10^{49}$~erg\,\cite{auriga05}. The LIGO collaboration also
published on the same giant flare, targeting times and frequencies
of the quasi-periodic oscillations in the flare's x-ray tail as well
as other frequencies in the detector's band, and setting upper limits on
GW energy as low as $8\times10^{46}$~erg for quasi-periodic signals
lasting tens of seconds\,\cite{matone07}.

\begin{scriptsize}
\begin{table*}
\caption{GW strain and energy upper limit estimates at 90\%
confidence ($\hrssn$ and $\egwn$) for the SGR 1806$-20$ giant flare
and the S5 SGR burst with the smallest limits on the ratio $\gamma =
\egwn/E_{\mathrm{EM}}$ for various circularly/linearly polarized RD
(RDC/RDL) and white noise burst (WNB) waveforms. Uncertainties
(given in superscripts for strain upper limits and explained in the
text) are folded into the final upper limit estimates. }
\begin{tabular}{@{\extracolsep{\fill}}lrlr|r|c||rlr|r|c}
 \hline \hline
 & \multicolumn{5}{c}{SGR 1806$-20$ Giant Flare} & \multicolumn{5}{c}{SGR 1806$-20$ GRB 060806 } \\
 Waveform type & \multicolumn{3}{c}{$ \hrssn [ 10^{-22}~ \rthz $] }  & $\egwn$ [erg] & $\gamma$  & \multicolumn{3}{c}{$ \hrssn [ 10^{-22} ~ \rthz $] }  & $\egwn$ [erg]  & $\gamma$  \\
 \hline
 WNB 11ms 100-200 Hz   & 22 & $^{ +1.3 ~ +5.6 ~ +1.2}$ &  $= 29$  $  $ & $\sci{7.3}{47}$ & $\sci{5}{1}$ & 3.4 & $^{ +0.0 ~ +0.4 ~ +0.2}$ &  $= 3.8 $ & $\sci{1.3}{46}$ & $\sci{4}{3}$ \\
 WNB 100ms 100-200 Hz   & 18 & $^{ +1.1 ~ +4.6 ~ +0.5}$ &  $= 24$  $  $ & $\sci{4.9}{47}$ & $\sci{3}{1}$ & 2.9 & $^{ +0.0 ~ +0.3 ~ +0.1}$ &  $= 3.3 $ & $\sci{9.1}{45}$ & $\sci{3}{3}$ \\
 WNB 11ms 100-1000 Hz   & 50 & $^{ +3.0 ~ +13 ~ +1.3}$ &  $= 66$  $  $ & $\sci{5.4}{49}$ & $\sci{3}{3}$ & 7.5 & $^{ +0.0 ~ +0.8 ~ +0.3}$ &  $= 8.3 $ & $\sci{8.3}{47}$ & $\sci{3}{5}$ \\
 WNB 100ms 100-1000 Hz   & 45 & $^{ +2.7 ~ +12 ~ +1.1}$ &  $= 59$  $  $ & $\sci{3.7}{49}$ & $\sci{2}{3}$ & 7.0 & $^{ +0.1 ~ +0.7 ~ +0.2}$ &  $= 7.9 $ & $\sci{6.8}{47}$ & $\sci{2}{5}$ \\
 RDC 200ms 1090 Hz   & 59&  $^{ +3.6 ~ +15 ~ +1.7}$ &  $= 78$  $  $ & $\sci{2.6}{50}$ & $\sci{2}{4}$ & 10 & $^{ +0.2 ~ +1.1 ~ +0.4}$ &  $= 12$  $  $ & $\sci{5.8}{48}$ & $\sci{2}{6}$ \\
 RDC 200ms 1590 Hz   & 93&  $^{ +5.6 ~ +24 ~ +2.8}$ &  $= 120$  $  $ & $\sci{1.4}{51}$ & $\sci{9}{4}$ & 15 & $^{ +0.6 ~ +1.5 ~ +0.5}$ &  $= 17$  $  $ & $\sci{2.5}{49}$ & $\sci{8}{6}$ \\
 RDC 200ms 2090 Hz   & 120&  $^{ +7.4 ~ +32 ~ +3.5}$ &  $= 160$  $  $ & $\sci{4.2}{51}$ & $\sci{3}{5}$ & 20 & $^{ +1.6 ~ +2.5 ~ +0.6}$ &  $= 24$  $  $ & $\sci{8.9}{49}$ & $\sci{3}{7}$ \\
 RDC 200ms 2590 Hz   & 150&  $^{ +9.1 ~ +39 ~ +4.1}$ &  $= 200$  $  $ & $\sci{9.8}{51}$ & $\sci{6}{5}$ & 24 & $^{ +3.1 ~ +3.0 ~ +0.9}$ &  $= 30$  $  $ & $\sci{2.2}{50}$ & $\sci{7}{7}$ \\
 RDL 200ms 1090 Hz   & 170&  $^{ +10 ~ +44 ~ +36}$ &  $= 240$  $  $ & $\sci{2.6}{51}$ & $\sci{2}{5}$ & 33 & $^{ +1.0 ~ +3.4 ~ +3.5}$ &  $= 38$  $  $ & $\sci{6.7}{49}$ & $\sci{2}{7}$ \\
 RDL 200ms 1590 Hz   & 260&  $^{ +16 ~ +68 ~ +32}$ &  $= 360$  $  $ & $\sci{1.2}{52}$ & $\sci{7}{5}$ & 44 & $^{ +2.2 ~ +4.6 ~ +6.3}$ &  $= 54$  $  $ & $\sci{2.8}{50}$ & $\sci{9}{7}$ \\
 RDL 200ms 2090 Hz   & 390&  $^{ +23 ~ +99 ~ +46}$ &  $= 520$  $  $ & $\sci{4.4}{52}$ & $\sci{3}{6}$ & 64 & $^{ +7.0 ~ +8.1 ~ +9.1}$ &  $= 83$  $  $ & $\sci{1.1}{51}$ & $\sci{4}{8}$ \\
 RDL 200ms 2590 Hz   & 440&  $^{ +26 ~ +110 ~ +63}$ &  $= 600$  $  $ & $\sci{8.9}{52}$ & $\sci{6}{6}$ & 79 & $^{ +10 ~ +10 ~ +9.7}$ &  $= 100$  $  $ & $\sci{2.6}{51}$ & $\sci{9}{8}$ \\
 \hline \label{table:bestresults}
 \end{tabular}
 \end{table*}
 \end{scriptsize}

In addition to the 2004 giant flare, the search described here
covers 190 lesser events which occurred during the LIGO S5 data
run.
Furthermore this search extends to the entire high sensitivity band
of the detectors, which makes it the first search sensitive to
neutron star $f$-modes, usually considered the most efficient GW
emitting modes\,\cite{andersson97}. Our upper limits on $\egw$
overlap the range of $\eem$ $10^{44}$--$10^{46}$~erg seen in SGR
giant flares\,\cite{hurley05, palmer05}. Most of the WNB limits, and
some of the RD limits, are below the $10^{49}$~erg maximum $\egw$
predicted in some theoretical models\,\cite{ioka01}. Our best upper
limits on $\gamma$ are within the theoretically predicted range
implied in\,\cite{ioka01}.

The Advanced LIGO detectors promise an improvement in $\hrss$ by more
than a factor of 10 over S5, corresponding to an improvement in
energy sensitivity (and therefore $\gamma$) by more than a factor of
100.  Thus within the next few years we expect to obtain GW energy
upper limits for the $f$-mode search that fall in the $\eem$ range
of giant flares, and for the unmodeled search that fall in the
$\eem$ range of intermediate bursts.

The authors are grateful to the Konus-Wind team and to S. Mereghetti
for information used in the S5 burst list, and to G. Lichti and D.
Smith for information on the giant flare event time. The authors
gratefully acknowledge the support of the United States National
Science Foundation for the construction and operation of the LIGO
Laboratory and the Science and Technology Facilities Council of the
United Kingdom, the Max-Planck-Society, and the State of
Niedersachsen/Germany for support of the construction and operation
of the GEO600 detector. The authors also gratefully acknowledge the
support of the research by these agencies and by the Australian
Research Council, the Council of Scientific and Industrial Research
of India, the Istituto Nazionale di Fisica Nucleare of Italy, the
Spanish Ministerio de Educaci\'{o}n y Ciencia, the Conselleria
d'Economia Hisenda i Innovaci\'{o} of the Govern de les Illes
Balears, the Royal Society, the Scottish Funding Council, the
Scottish Universities Physics Alliance, The National Aeronautics and
Space Administration, the Carnegie Trust, the Leverhulme Trust, the
David and Lucile Packard Foundation, the Research Corporation, and
the Alfred P. Sloan Foundation. K.C.H. is grateful for support under
JPL Contracts 1282043 and 1268385, and NASA grants NAG5-11451 and
NNG04GM50G. This paper is LIGO-P070105-03-Z.

\end{document}